\documentclass{article}
\usepackage{arxiv}
\usepackage[utf8]{inputenc} 
\usepackage[T1]{fontenc}    
\usepackage{hyperref}       
\usepackage{url}            
\usepackage{booktabs}       
\usepackage{nicefrac}       
\usepackage{microtype}      
\usepackage{lipsum}

\usepackage{cite}
\usepackage{amsmath,amssymb,amsfonts}
\usepackage{pifont}
\usepackage{multirow}
\usepackage{algorithmic}
\usepackage{graphicx}
\usepackage{textcomp}
\def\BibTeX{{\rm B\kern-.05em{\sc i\kern-.025em b}\kern-.08em
    T\kern-.1667em\lower.7ex\hbox{E}\kern-.125emX}}
    
\title{User Incentives for Blockchain-based Data Sharing Platforms}

\author{
  Vikas Jaiman \\
  Institute of Data Science \\
  Maastricht University \\
   6211 LK Maastricht, The Netherlands \\
  \texttt{v.jaiman@maastrichtuniversity.nl} \\
  \And
  Leonard Pernice \\
  School of Business and Economics \\
  Maastricht University\\
  6211 LK Maastricht, The Netherlands \\
    \texttt{l.pernice@alumni.maastrichtuniversity.nl} \\
   \And
 Visara Urovi \\
  Institute of Data Science \\
  Maastricht University\\
  6211 LK Maastricht, The Netherlands \\
  \texttt{v.urovi@maastrichtuniversity.nl} \\
}

\begin{document}
\maketitle

\begin{abstract}
Data sharing is very important for accelerating scientific research, business innovations, and for informing individuals. Yet, concerns over data privacy, cost, and lack of secure data-sharing solutions have prevented data owners from sharing data. To overcome these issues, several research works have proposed blockchain-based data-sharing solutions for their ability to add transparency and control to the data-sharing process. Yet, while models for decentralized data sharing exist, how to incentivize these structures to enable data sharing at scale remains largely unexplored. In this paper, we propose incentive mechanisms for decentralized data-sharing platforms. We use smart contracts to automate different payment options between data owners and data requesters. We discuss multiple cost pricing scenarios for data owners to monetize their data. Moreover, we simulate the incentive mechanisms on a blockchain-based data-sharing platform. The evaluation of our simulation indicates that a cost compensation model for the data owner can rapidly cover the cost of data sharing and balance the overall incentives for all the actors in the platform.
\end{abstract}


\section{Introduction}
\label{sec:introduction}
Today, large amounts of data are being generated every second, yet data remains siloed in the databases of hospitals, companies, and research institutions around the globe. Data sharing is known to accelerate scientific research, improve business innovations, and to inform decision-making~\cite{Borgman2012,Nelson2009,OpenData2017, tenopir2011sharing, european2007shared}. 
Yet, several factors contribute to the lack of data sharing in practice~\cite{Borgman2012,OpenData2017} including legislation, institutional concerns, task complexity, use and participation, information quality, and technical concerns~\cite{janssen2012benefits}. Stringent data protection laws impede the procurement of large amounts of data. Regulations such as the General Data Protection Regulation (GDPR)~\cite{GDPR}, Federal Trade Commission (FTC) Act~\cite{ftc}, California Consumer Privacy Act 2018 (CCPA)~\cite{ccpa}, UK Data Protection Act 2018~\cite{uk}, and Australia Privacy Act 1988~\cite{au} determine several mechanisms of data protection, amongst which GDPR is explicit on the rights of individuals to request information on collected data, the purpose of use, with whom data is shared, as well as request rectification or deletion of their data. 
At a minimum, sharing personal records (i.e. patient data) requires involving individuals in the data-sharing process. This process includes the expression of informed consent which is normally collected via consent forms specifically tailored for a given study. Secondary use of data is possible when the original consent and the secondary use are compatible. This process is highly inefficient, often infeasible, and contributes to the lack of secondary data re-use~\cite{consent}. The typical solution to these issues has been third-party involvement. Users consent to share their data with third parties by agreeing to license agreements, without necessarily being aware of the connected potential risks or consequences. With this approach, 
data is collected and curated by different organizations or companies, to often be sold to those who can afford to pay the price. An example of this is the deal the direct-to-consumer genetics company 23andMe made with the biopharmaceutical giant Genentech, to sell access to genomic data on roughly 3000 Parkinson’s patients for a total of 60 million dollars~\cite{23andMeDeal}. 
Data collections can then become a source of large profits for data processors, however, the individuals whose data are being processed are rarely involved or compensated in this process. Decentralized data-sharing networks (such as blockchains) overcome many data-sharing issues by adding a transparency layer for all data transactions and by directly enabling participants to control their own records~\cite{PRIMABase}. Current decentralization techniques however are not costless. Most open blockchain platforms are based on payments per transaction, thereby, a transparency layer and individual data control come with costs and with importunities for building appropriate incentives for a data market to emerge.

Previous research~\cite{DataSharingIncentives, ClinicalDataSharingIncentives, incentiveprivacy} examine incentives that motivate data-sharing and establish several key benefits that include verification of previous research, new interpretations, improved data integrity, resource optimization, guard against falsification, and facilitation of researcher training.
However, state-of-the-art blockchain platforms~\cite{MedRecBase, OceanMarketplace, DataSharingFramework} don't explore the incentives structures for data providers. To date, there is a lack of data-sharing infrastructure that can sustainably facilitate the incentive structures to motivate data sharing on the blockchain.

\noindent\textbf{Contributions.} To overcome these limitations, in this paper, (i) we develop an incentive model to motivate user participation on a blockchain-based data sharing platform, (ii) we create a functioning prototype from the resulting incentive model, and (iii) finally we conduct extensive experiments and analyze the solution in scenarios simulating real-life user interactions. Specifically, the scenarios showcase the accruing operational costs inherent to the blockchain implementation. By conducting these simulations that incorporate user incentives, we create a foundation that showcases the boundary conditions of a blockchain-based solution for data-sharing. Our approach shows which costs can be expected from a sophisticated data-sharing platform based on blockchain technology, and how they can be covered by using such platform. 
The novelty of this work is an analysis of cost-benefits for two main incentive systems, i) sharing the costs of data providers and, ii) profit-making for data providers. 
Our simulation indicates that a cost compensation model for data provider quickly covers the cost of data sharing. 

The remainder of this paper is structured as follows. Section~\ref{sec:background} discusses the background work related to blockchain platforms. Further, section~\ref{sec:architecture} presents the architecture of the incentive model. Next, section~\ref{sec:implementation} presents the implementation of our solution. Section~\ref{sec:evaluation} discusses the evaluation of the proposed model followed by a discussion in section~\ref{sec:discussion}. Section~\ref{sec:relatedwork} discusses related work and highlights the limitations of the state-of-the-art. Finally, section~\ref{sec:conclusion} concludes the paper and presents future works.
\section{Background}\label{sec:background}
In this section, we explain Ethereum blockchain, incentives, and our baseline data sharing platform. 
\subsection{Blockchain as a decentralized network}\label{subsec:blockchain}
Blockchain is a decentralized network of nodes that maintains a shared ledger of transactions. Blockchains consist of chained transaction blocks that are validated and added to the blockchain by the nodes in the network. In order to add new blocks, they are concatenated with the last confirmed block in the ledger. The cryptographic hash of the previous block is added to the newly created block to generate an updated hash value. Once the block is added, the transactions contained in the block are permanent and immutable. 
Blockchains use validation nodes, also called \textit{miners} to update the ledger. The validation is pre-established by a \textit{consensus mechanism} that specifies what determines a valid block. Consensus mechanisms differ however they all focus on rewarding validators for maintaining the state of the blockchain. Open blockchains do not limit or control the validation nodes, however, becoming a validator requires sometimes substantial initial investments, thus it cannot be seen as a main incentive mechanism for data-sharing platforms because not all users will become validators.
We focus on openly accessible decentralized networks, such as Ethereum \cite{Ethereum2.0} for the openly accessible ledger and the general-purpose architecture. Using the Ethereum blockchain as the starting model for monitoring data transactions, individuals are able to inspect and control their data sharing preferences. Importantly, blockchain technology circumvents the need for centralizing data into a third party and supports open data-sharing agreements that are validated by the network. Blockchain networks, by design, introduce \textit{transaction costs}. This cost is a computational cost which in Ethereum is measured in \textit{gas\footnote{a measure of the computational effort required to perform an operation}}. Gas is attributed an Ether value, measured in Wei\footnote{Wei is the smallest denomination of Ether.}. Two additional open and general purpose blockchain platforms have been recently launched, Cardano\cite{cardano} and Polkadot\cite{polkadot}.
We choose Ethereum for its well-established platform, yet the overall findings of this work can be applied to any of these platforms by accounting for the transaction fees and the computational costs of these other networks.

\subsubsection{Smart Contract}
A smart contract is a digital protocol that facilitates, verifies, and executes one or multiple transactions~\cite{smartcontracts}. Smart contracts, similarly to real-life physical contracts, translate contractual clauses between two parties. They achieve this with rules that are written into executable code. Smart contracts are executed independently by the network nodes and become immutable after deployment. Ethereum smart contracts provide a generic mechanism for building applications that require agreements between two or more parties. Using smart contracts, the transactions become valid only when the contractual agreement are met, resulting in the storage of the transaction in the blockchain. We use smart contracts to define data-sharing and incentive rules between the data providers and data requesters.  

\subsubsection{Tokens}
Ethereum tokens are a special sub-type of cryptocurrency, usually defined as fungible, exchangeable assets. They are created from specialized smart contracts and are mostly used to create secondary economies on top of the Ethereum network. A noteworthy example of this is the DAI stablecoin~\cite{DAI}, which is based on the ERC-20 token standard~\cite{ERC20} and is perfectly robust against the volatility that other cryptocurrencies such as Ether or Bitcoin commonly experience. The Ethereum Improvement Proposals (EIP)~\cite{EIP} is a collection of standards, new features for the Ethereum network. The main advantage of tokens is a platform-wide standard practice for method definition which leads to fewer faulty contracts and easy implementation of interoperability. We use tokens for access control to data, thus providing exclusive data access based on the established agreements between data-provider and data requesters. More specifically, tokens provide a way to link the blockchain irrefutable transactions with data access control, in a way that data is not accessible to other users (i.e. data requesters) unless there was a prior agreement reached within a smart contract.
ERC-20 is a standard API for tokens in smart contracts that provides base functionality to transfer tokens or approval for third parties to transfer tokens. Today, there is no mechanism to protect against faulty token transactions, making them irrecoverable in certain cases. 
ERC-721 is based on ERC-20 and implements a token standard where each token is unique and can have different values (non-fungible). This makes it useful for representing physical property and other such assets. ERC-721 tracks ownership of each token individually. Additionally, tokens can be deleted and associated methods are robust against faulty inputs. However, it does not provide any type of data structure to associate tokens with individual properties.
In this paper, we adapt the ERC-721 token standard to represent a unique access key to specific datasets, since it is the closest standard to our token implementation. 

\subsection{LUCE}\label{subsec:luce}
LUCE~\cite{LUCEBase} is a blockchain-based data-sharing platform that allows data providers to verify for which purpose, by whom, and in which time-frame their data is used. LUCE allows users to share and reuse data in compliance with license agreements. LUCE ensures compliance with the GDPR by giving the data provider personalized methods to control their data. Additionally, the data provider can issue updates, change the required license, or completely delete the dataset. All of these changes perpetuate through the system. A data provider can generally publish and update their datasets. When a dataset is published, the data provider provides information on it. This includes meta-information on the dataset, access requirements, and an access link. Afterwards, this information is saved to the respective smart contract. Thus, each dataset is connected to a separate smart contract. This allows the provider intricate control over how each dataset should be accessed by requesters. If a data requester fulfills the requirements set by the data provider, they can make access requests, which are time-bounded. However, requesters can also renew their access time. The smart contract provides GDPR compliance, which binds all requesters to the access conditions of each respective dataset. The supervisory authority (e.g. governmental institution) is responsible for enforcing the rights of the data subjects and general prevention of abuse of the platform. If there is a legal issue, i.e. a data requester's non-compliance with the license agreement of a specific dataset, the supervisory authority is responsible for auditing the related metadata and various system interactions of the parties involved. 

Using LUCE as a basis for a decentralized sharing network, we extend the model with an incentive model and analyze the scenarios simulating real-life experiences. In our approach, we showcase the accruing operational cost for data sharing on the LUCE platform. 
\subsection{Incentives}
For data providers, compensations can be monetary or reputation-based. However, there are also aspects of the system that may disincentivize data providers, namely, unavoidable costs arising from the usage of the platform. The requirement from the perspective of data requesters for using the platform is data availability, which is provided by incentivizing data providers to use the platform. There are several types of important incentive mechanisms to consider in decentralized networks:
\subsubsection{Research} Data requesters' are intrinsically motivated to use data-sharing platforms due to the value of data in research. This ties into the general main incentive of the platform, which is promoting data-sharing on a large scale. This incentive is powerful for all involved parties (data requesters and providers) due to the potential results from research on shared data (for example medical research data). Data providers may be interested in findings but also might simply regard data-sharing as a goodwill act towards society.
\subsubsection{Monetary} Monetary incentives in decentralized networks are important to consider, especially for data providers. Decentralized networks distribute operational costs, which implies that a data provider will incur initial costs to share data and to keep them up-to-date. Monetary incentives may be an incentive for data providers. Data requesters on the other hand, maybe willing to pay for data access.
\subsubsection{Reputation} An incentive that does not directly involve monetary incentives is reputation~\cite{blockchainreputation,blockchainreputation2}. Data providers may share data on the platform to receive mentions and recognition for data re-use. This is particularly relevant to researchers who become data providers to share their data collections for further re-use.
\subsubsection{Knowledge} The most important type of incentive will be created by the knowledge shared by data requesters. This could be in the form of analytical models, which, if returned to data providers, can provide a personalized outcome for every data provider.

In this paper, we focus on monetary incentives as these are the incentives that we can realistically simulate, without extensive surveys and practical experimentation in a real-world test environment. Moreover, monetary compensation and \textit{cost} allocation are the first elements to address in decentralized data-sharing networks as the occurring costs can discourage data providers from participating in data-sharing.
\section{Incentive model architecture}\label{sec:architecture}
\begin{figure}[ht]
\centering
\includegraphics[width=\linewidth]{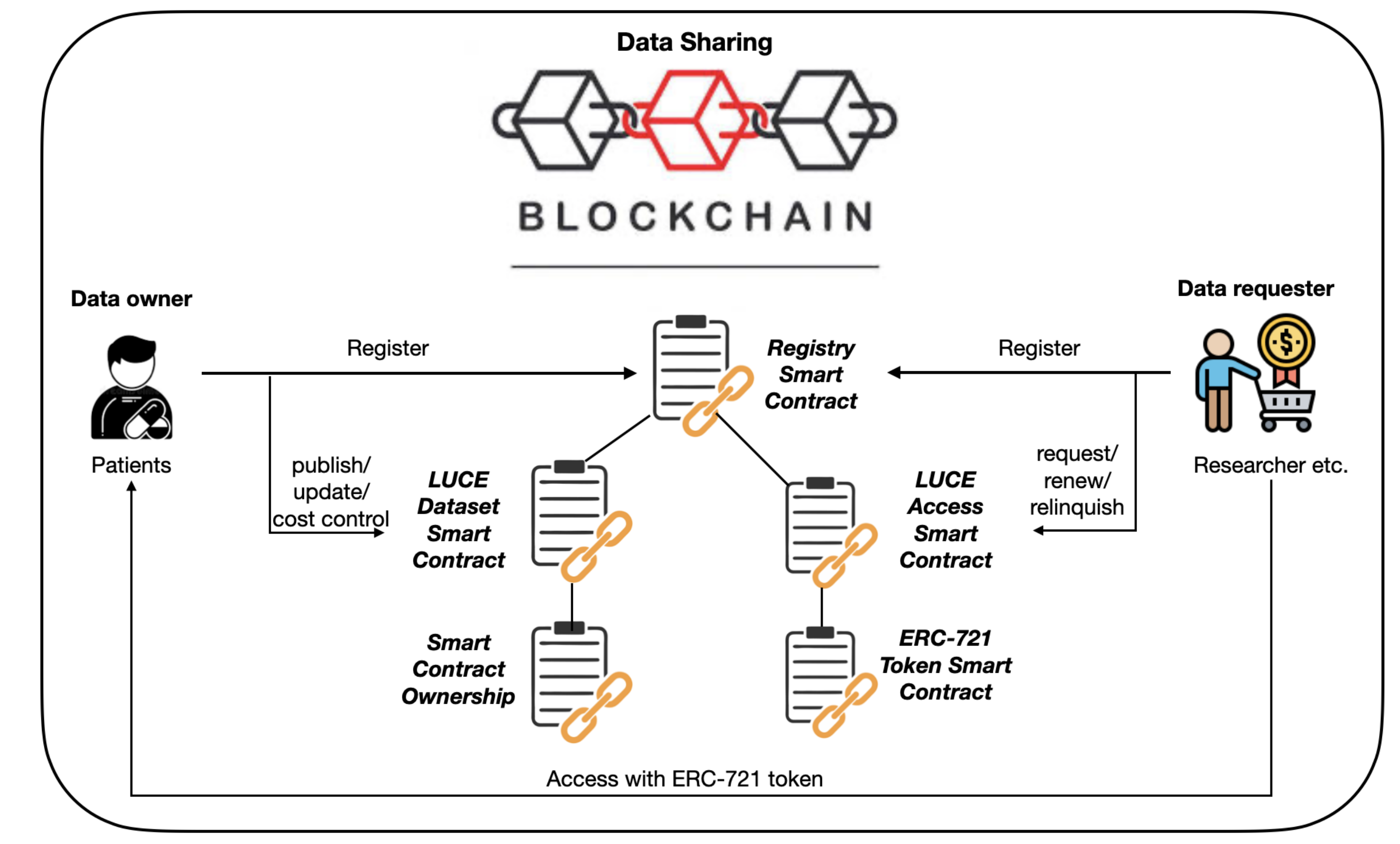}
\caption{User incentives architecture on LUCE}
\label{fig:LUCE Overview}
\end{figure}
Figure \ref{fig:LUCE Overview} shows the architecture of the user incentive model on LUCE. We develop incentive-based smart contracts to have interactions between the data providers and data requesters. We present the descriptions as follows:  
\begin{enumerate}
    \item \textit{Registry smart contract} - provides authorization for data publishing and access requests.
    \item \textit{Dataset smart contract} - handles data publishing, updates, and cost control.
    \item \textit{Smart contract ownership} - defines the connected contracts as owned by the data provider that deploys the main contract and is connected to an additional module that allows the owner to delete their smart contract.
    \item \textit{Access smart contract} - handles access and access renewal requests by data requesters and is connected to the ERC-721 token generation contract.
    \item \textit{ERC-721 Smart Contract} - adapted token standard that handles the token logic that is key to accessing the data.
\end{enumerate}

\subsection{Registry Smart Contract}
\label{LUCE Registry Smart Contract}
 We develop a global registry smart contract that can interface with the LUCE smart contracts to provide access exclusivity to particular individuals. This registry is deployed and controlled by the institution responsible for verifying a registrant's information. When a user registers, their information is connected to a wallet in the blockchain, i.e. they are anonymous, yet unambiguously associated with their valid license information. Thus, a user's public key is synonymous with their identity, and, since it is impossible to deduce the identity of the owner from a public key, they can act anonymously. The only information associated with these public keys is the requester's license or the provider's publishing permission, and the only parties privy to identifying information are the owner of the key and the authority that verified the owner's identity. When an individual makes their first transaction on the blockchain, e.g. publishing a dataset or requesting access to a dataset, their registration information is verified internally. This ensures that no unauthorized individual can interact with the relevant smart contracts, even if they possess the knowledge to circumvent the LUCE.

However, this centralized control structure functions only as a gateway to the platform and has no influence on the actual data-sharing process, any possible monetary transactions, or even any purview of how the platform is used.


\subsection{Dataset Smart Contract}
\label{Dataset Smart Contract}
The dataset smart contract establishes control for the data provider over their dataset. Each dataset must be published on a separate dataset smart contract. This provides the supervisory authority with the structure of a comprehensive record, and meta-information can be changed via an update. Due to GDPR requirements, each update that results in a change in the meta-information of the respective dataset requires all active data requesters to confirm their compliance. Specifically, they will be notified of the update, and until they have updated their own copy of the data and confirmed this via a special compliance function, the respective requester cannot make access requests to the data. A different type of update is if the data provider changes the required license to access the data. In this case, all tokens with the wrong license type will be deleted by the system, and data requesters get notified. All affected data requesters must then confirm their compliance with this change, and delete their copy of the dataset. Finally, the data provider can establish how the contract handles arising costs. 

\begin{enumerate}
    \item \textit{Scenario 1}. No compensation - each party pays only their own arising costs.
    \item \textit{Scenario 2}. Cost compensation - the data provider's costs are covered by the data requesters.
    \item \textit{Scenario 3}. Profit - the data provider seeks to profit from sharing their data.
\end{enumerate}

Generally, the scenarios are meant to showcase how the system reacts to different incentives being implemented. In general, scenario 1 represents no incentives apart from those naturally arising from using the system, meaning data providers are most likely disincentivized from using the system since they incur costs by using it. Scenario 2 seeks to remedy this by implementing a structure that asks data requesters to pay a fraction of the provider's total running costs at the time of their request. This results in a gradual decline in running costs for the provider, which represents a fairness consideration. Therefore early data requesters will pay relatively more than later data requesters since already transmitted fees are deducted from the running cost in the smart contract. Finally, scenario 3 shows how profits may be generated, and how soon the break-even point is reached. 

To test these scenarios, the dataset smart contract allows data providers to manipulate settings regarding cost allocation. Data providers can set a percentage profit margin that describes the total earnings aimed for it. 

\subsection{Smart Contract Ownership}
\label{Owned and Destructible Smart Contract}
This module establishes a method to control which individuals (i.e. public addresses) can call certain core functions of the underlying contracts, such as issuing an update to the data. When a data provider deploys their copy of the template smart contract to publish a dataset, their address is immediately noted as the owner of that smart contract, and all smart contracts that inherit it. The most important function needs authorization of the owner is the destruction of the contract and all super and subordinate contracts. This function is implemented in a smart contract sub-module, which allows the owner to send all funds from the internal balance of the smart contracts to their public address while setting all internal variables to zero. Therefore any subsequent call to this contract will be voided. With this, we implement the data providers' right to delete their data (GDPR, Article 17~\cite{GDPR}). However, it is important to make sure requesters are adequately informed of this change since they could otherwise mistakenly transfer funds to the destroyed contract, which would result in those funds being lost forever. LUCE automatically delists a deleted dataset's contract address from the data catalog. 

\subsection{ERC-721 Smart Contract}
\label{sec:ERC721SmartContract}
The purpose of generating tokens as access keys to datasets is that they represent a fixed, standardized data structure that can be easily interfaced. For this, the token must supply several properties: It must be unique, provide adequate control methods and internal data structures, and be easily traceable. The ERC-721 smart contract module establishes a list of all tokens generated. Factually, a token is simply an entry in this list, represented by a unique ID that unambiguously identifies it. This ID is associated with an \textit{owner}, i.e. the individual (public address) that minted it. Only the owner can transfer the token to another individual. The transfer of a token results in all associated values being accessible and controlled by the new owner. Since requesters should not have the ability to transfer their token to other requesters, therefore, we created a new structure that associates the token ID with its \textit{user}, i.e. requester. This results in the \textit{user} of a token only having limited control over it, i.e. they can use it for three purposes: accessing the data, renewing access time to the data, and deleting their access to the data. Moreover, we created an internal data storage structure that saves meta-information on the requester and the token (e.g. license, access time, etc.), which only the data provider, respective data requester, and supervisory authority can access. By limiting access to this information we protect the privacy of the data requester.


\subsection{Access Smart Contract}
\label{sec:LUCESmartContract}
\begin{figure}[ht]
\centering
\includegraphics[width=\linewidth]{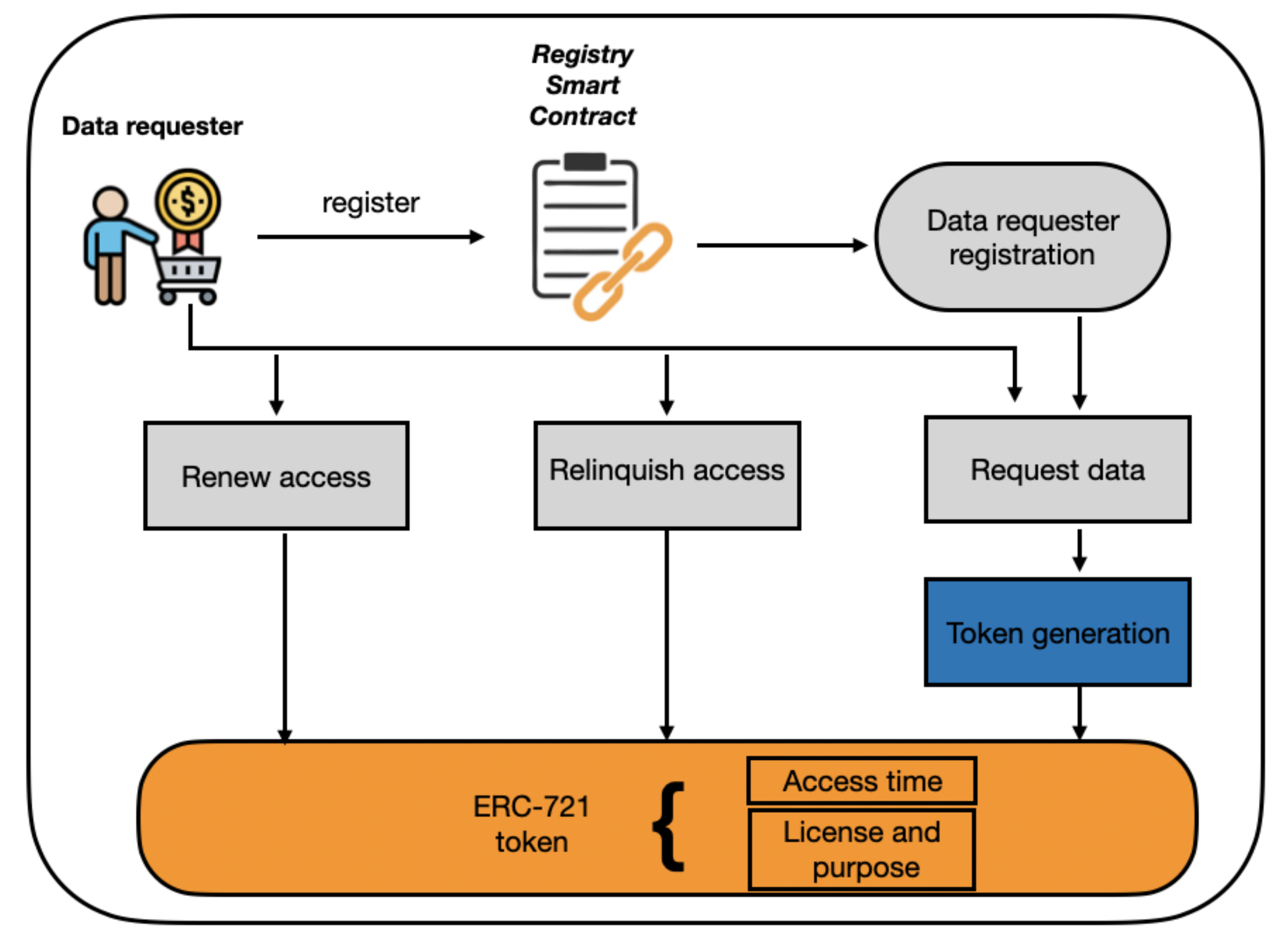}
\caption[LUCE Access Module]{Access methods for a data requester in the access smart contract module.}
\label{fig:Requester Functions}
\end{figure}

This contract holds the methods for data access and access renewal requests, implements cost coverage and GDPR compliance systems, and allows data requesters to relinquish their access if it is no longer needed. Whenever a data requester makes an access request, this contract establishes a connection with the LUCE registry to confirm their license. In addition, we also implement the cost coverage system, which applies to the settings controlled by the data provider. If all access requirements are met, the contract will generate a unique token via the ERC-721 contract~\cite{ERC721}. This unique token serves as an access key for the data requester to the data. Figure \ref{fig:Requester Functions} shows an overview of the methods data requesters have at their disposal.
When the data requester successfully gains access to the data, by default they are granted two weeks of access time, after which they must either actively delete their copy of and access to the data, or renew their access time. We implement methods for both options. Access time renewal necessitates that the data requester has actively confirmed their compliance with GDPR requirements following a potential update by the data provider. The compliance function signifies that the requester that calls it has actively confirmed their compliance with all past updates. This serves as a marker for the supervisory authority should there ever be a complaint against the respective data requester that requires investigation. If this requirement is fulfilled, the data requester is given more access time. Finally, if the data provider wishes to relinquish their access to the dataset, they can do so by disassociating their public address (i.e. anonymized identity) with the token. This causes the respective data requester to lose access to the data unless they decide to make a new access request.

\section{Implementation}\label{sec:implementation}
In this section, we provide the implementation details of the smart contracts and the extension of LUCE~\cite{LUCEBase} with the user incentive model proposed in this paper.

\noindent\textbf{Experimental setup.} We implement the smart contracts of the incentive model in Solidity~\cite{soliditybackground}, a language for smart contracts provided by Ethereum. Our incentive model is then embedded into the LUCE platform -- a blockchain-based data sharing platform deployed on Ganache Ethereum network~\cite{Ganache}. To run our experiments, we use LuceDocker -- a dockerized version of LUCE. The dockerized image is deployed on a server hosted by the Institute of data science, Maastricht University, and the web-hosted version is accessible via \url{https://luce.137.120.31.102.nip.io}. Moreover, LuceVM virtual machine~\cite{LUCEBase} is also available to run the experiments which is running on a 64 bit Ubuntu 16.04 LTS (Xenial Xerus) Linux operating system. The virtual machine is equipped with 1024 MB RAM. Our incentive model implementation is available as open-source\footnote{https://github.com/vjaiman/LUCE\_Incentives}. 
 
Our incentive model is implemented on top of the Ethereum blockchain. It uses Web3 javascript libraries~\cite{web3js} to interact with the Ethereum blockchain. It uses Django~\cite{django} for implementing the user interface. The data providers interact via the Django web framework to share the data and specify the associated incentives. It stores the link between the smart contract and the corresponding datastore location. Through the LUCE platform, the model interacts with Ganache~\cite{Ganache}, a test network that creates a virtual Ethereum blockchain and generates pre-configured accounts that are used for the development and testing. The accounts are pre-funded which enables the deployment of the contracts. Ganache provides the balance in \textit{ether} and notifies the \textit{gas} used for running the transactions.

\subsection{Data provider cost allocation control}\label{Cost allocation control}
In our incentive model, the running costs after a transaction are equal to the running costs before a transaction in addition to the cost of the transaction times the \textit{profit margin}. 
\begin{center}
    $currentCost_t = currentCost_{t-1} + gasUsed * gasPrice * \frac{profitMargin}{100}$
\end{center}
The \textit{profit margin} describes the total earnings aimed for, expenses and returns, and can be set via the \textit{setProfitMargin} function. If a data provider doesn't want to make a profit, it is equal to $100\%$ i.e. 100\% of the pure costs of the data provider. If a data provider wishes to generate profits from sharing their data, they must declare their desired earnings as a linear combination of their costs. In addition,  by calling the \textit{setMultis} function, the data provider can control the percentage of the running costs that each data requester must pay upon access or access time renewal request.
The \textit{providerGasCost} modifier applies regardless of the running scenario and represents a convenient way for the data provider to keep track of their running costs in all scenarios. By using this modifier to measure costs arising from publishing data, we essentially ask the data provider to make an initial investment. This is beneficial for several reasons. First, it discourages poor quality data from being shared. Second, it reduces the complexity of the system by a large margin, since the alternative is employing meta transactions\footnote{Meta transactions are a special type of transaction that is signed by one individual and then published so that an arbitrary different individual can execute them in the name of the signer \textit{MetaTransactions}.}, which would allow the data provider to sign a prepared transaction. Afterwards, the data requester transacts the data provider's signed transaction to the blockchain and thus pay the associated gas cost directly. 

\subsection{Data requester methods} \label{subsec:datarequestermethods}
In this section, we explain the technical implementation of the core functionalities of the smart contracts used in our incentive model.
\subsubsection{Request access} 
In our incentive model, access rights are distributed via tokens, which are associated with the data requester once their legitimate claim has been verified. To do this, a data requester has to follow a range of requirements; i) a dataset must be published, ii) the requester must not yet own an access token to this dataset, iii) the requester must be registered and possess the same license as is required for accessing the data, and iv) finally, smart contract checks for which scenario it is running. If it is scenario 2 or 3, a requester must submit an appropriate amount with their access request.
Once the data requester receives an access token, they can call the \textit{getLink} function to download the dataset.
\subsubsection{Renew access time}
\label{subsec:renewaccesstime}
The access time associated with any access token is fixed to a reasonable amount of time (e.g. 2 weeks). If a data requester needs the data access for longer, it can renew the access time. For this, a data requester must have an access token to that specific dataset. Second, it must have confirmed compliance with any previous updates. The \textit{confirmCompliance} function allows data requesters to notify the system of their GDPR compliance following an update, which allows them to renew their access time to the data. 

%

\subsection{Relinquish access}
\label{relinquishaccess}
The data requester with a token has a limited range of actions they can take, the most relevant of which are accessing the data, renewing their access time to the data, and deleting their token should that ever be required. To delete their token, a data requester must call the \textit{burn} function, or the smart contract calls it upon a change in the license requirement. When this requirement is fulfilled, the function first notes the remaining access time (0 if the access time is expired). Then, the internal \textit{\_burn} function of the ERC-721 token standard is called, which associates the token with the null address i.e. it can no longer be used. Regardless of how the function is called, the data requester is notified of the event. If the token deletion was issued by the data requester, their compliance is set to \textit{true} since token deletion should always involve the deletion of the requester's copy of the dataset as well. If the token deletion was issued by a change in the license type, compliance is set to \textit{false}.
\section{Evaluation}\label{sec:evaluation}
In this section, we evaluate the effectiveness of \textit{monetary} incentives. Our evaluation aims at answering the following questions:
\begin{enumerate}
    \item How do costs arise over time from using the system?
    \item How long does it take to cover the costs in scenarios 2 and 3?
    \item How to find a balance between cost coverage for the data provider and fair payment amounts for all data requesters?
\end{enumerate}

\subsection{Initialization}
\label{Initialization}
We use Ganache~\cite{Ganache} to generate 1000 accounts which are prefunded with 100 Ether. Gas consumption varies based on the complexity of the functions defined in the smart contract.  We consider the gas price of 72 Gwei according to the current date\footnote{22/06/2021} with corresponding Ether price (1 ETH == \$1716.52)~\cite{Ethgasstation}.  Our simulation runs each iteration of the loop which signifies the passing of 1 \textit{period}. In each \textit{period} multiple \textit{actions} can be made. An \textit{action} in this context refers to one of four possible decisions being made: \textit{publishing data}, \textit{updating data}, \textit{requesting access}, or \textit{renewing access time}. Each potential data provider and data requester is associated with a certain probability of taking \textit{action}. We make the assumption that the chance of data requester taking \textit{action} underlies normal distribution parameters with independent, identically distributed variables, since this is the most commonly occurring distribution in nature:

\begin{center}
    $X \sim (\mu, \sigma^2)$
\end{center}

For simplicity's sake, we center our distribution around 0 (\(\mu = 0\)) and assume standard deviation is 0.1 (\(\sigma = 0.1\)). To associate each account with a normally distributed probability, we first generate 1000 random values of a normal distribution with the aforementioned parameters. Since the resulting values do not lie between 0 and 1, we normalize them. This results in a vector of random, normally distributed probabilities, which we append to the user accounts list. Thus, a data requester will, on average, have a 50\% probability to make an access request in a period. However, since we do not expect data requesters to require access to a specific dataset for an indefinite amount of time, we adjust their probability of taking \textit{action} downwards by a factor of 0.75 each time after they renew their access time to the data. This results in data requesters renewing their access time only very rarely after the fifteenth time (corresponds to 0.5*0.75$^{15}$=0.668\%). Thus we achieve a natural balance of data requesters starting, continuing, and stopping to renew their respective access time and avoid exponential growth of \textit{actions} being taken per \textit{period}, which would be highly unrealistic. We do not simulate data requesters burning their tokens at that point, since it is irrelevant for the data provider's costs. 

For data providers, we assume that the probability of choosing to publish is far lower than for an average data requester making an access request. Therefore, each data provider is given a uniformly distributed probability to publish that lies between 1\% and a maximum probability specified by us (default is 5\%). This overwrites the normally distributed probability assigned to the Ganache accounts designated as data providers. This reflects our assumption that data providers are generally less numerous than data requesters and would thus take \textit{action} less often.

\noindent\textbf{Assumptions.} 
We make the following assumptions about data providers and data requesters for the simulations.
\begin{itemize}
    \item The probability of a data provider deciding to publish their dataset is lower than the probability to update it after publishing. 
    \item The probability of both publishing and updating a dataset is constant, independent of consequent potential costs arising, and independent of the number of data requesters who have access to the dataset. 
    \item The probability of publishing is independent of the type of dataset. 
    \item The probability of data requesters taking \textit{action} decreases over time. Therefore, no data requester will continue to renew access to a single dataset indefinitely. 
    \item Data requesters have an unlimited amount of money potentially available to request access or renew access time to datasets. 
\end{itemize}

\subsection{Starting the simulation}
\label{Starting the simulation}
The first action in each simulation instance is the first data provider publishing their dataset. 
In each \textit{period} we check for each of the four possible \textit{actions}:
\begin{itemize}
    \item \textit{Publish}: exactly 1 data provider has the chance to publish (denoted by their probability of taking \textit{action}). Until they do publish, no other data provider will be able to publish. This represents the passage of time (\textit{periods}) between different providers publishing their data. 
    \item \textit{Update}: each data provider with a published dataset has the chance to issue an update. We assume that a data provider, once they published their dataset, is legally required to update it regularly, and we increase the chance to update by a certain factor.
    \item \textit{Request}: exactly 1 data requester has the chance to request access to a randomly determined dataset among those available. If this data requester does not request access, they will have the same chance to do so in the next iteration of the loop until such a time where they do make the decision to request access. Afterwards, the next data requester in line has the chance to make a request. This simulates the potential time gap between different requesters making access requests. 
    \item \textit{Renew}: each data requester with an access token will have the chance to renew their access time to the data. In our simulation, we assume that requesters will only renew access time if it has expired since this is economical behavior. A data requester may not know precisely for how long they need access, thus it makes sense to add access time only when needed, especially since potential costs in scenarios 2 and 3 are likely to be lower with each passing \textit{period}. 
\end{itemize}

We simulate the passage of time by assigning probabilities to users that might or might not take \textit{action}. On the other hand, we attribute access times in real seconds to the tokens generated upon a successful request or access renewal. Since the simulation would be flawed if these two systems do not operate synchronously, we implemented a condition that disallows access time renewal until 2 \textit{periods} after the requester's last \textit{action}. This reflects the idea that a \textit{period} is roughly equivalent to a week, thus each data requester would be able to renew their access to the data for two weeks.


\begin{table}
\centering
\caption{Parameters used in the incentive model.}
\label{table:parameters}
\setlength{\tabcolsep}{3pt}
\def\arraystretch{1.5}
\begin{tabular}{p{110pt}p{240pt}}
 Actions & Representation  \\\hline 
\textit{actionTicker} & represents number of \textit{actions}. This controls the length of simulations.  \\
\textit{maxDataProviders = 1} & represents the default maximum number of data providers we allow in this simulation.  \\
\textit{providerProbabilities = 0.05} & represents the default maximum probability of a new data provider publishing or updating their data. \\
\textit{updateMultiplier = 5} & represents the default multiplier that increases the chance of a data provider making an update to their data.\\
\hline
\end{tabular}
\end{table}
\begin{table}
\centering
\caption{Cost parameters used in the incentive model.}
\label{table:costparameters}
\setlength{\tabcolsep}{3pt}
\def\arraystretch{1.5}
\begin{tabular}{p{110pt}p{240pt}}
Parameter & Representation  \\\hline 
\textit{totalCost} & a running total of all arising costs, regardless of how or where they arise. \\ 
\textit{transactionCost} & the total cost of the transaction resulting from the user's \textit{action}. \\ 

\textit{currentExpectedCost} & the expected cost for a data requester before they make a transaction\\ 
\textit{nextExpectedCost} & the expected cost for a data requester after they make a transaction.  \\ 
\textit{providerEarnings} & a running total of the amount transmitted to the contract as payment.\\ 
\textit{providerCost} & a running total of the costs arising from the provider taking \textit{action} (i.e. publishing or updating their data).  \\
\hline
\end{tabular}
\end{table}
\subsection{Determining optimal parameters}
\label{Determining optimal parameters}
As seen in Table~\ref{table:parameters}, the most pivotal variables (apart from the scenario itself) are the \textit{actionTicker}, and the cost fraction data requesters must pay when making \textit{access requests} or \textit{renewing} their access time. We simulate scenario 2 to determine the optimal values for these variables since this is the most dependent on actions. We observe that a high percentage cost distribution (i.e. the fraction a data requester must pay in return for access) leads to a too rapid decline in the running contract cost and immediate coverage of new arising costs whenever the data provider updates. It is inherently unfair to the data requesters since some will pay high amounts while others pay almost nothing. On the other extreme, when data requesters pay only a small fraction of the running contract costs we observe a balancing of revenue and expenses above zero, which is not the goal of scenario 2. Thus, we conclude that the fraction must lie between the extremes to be effective i.e. 5\% cost coverage and 500 \textit{actions}. 
The \textit{profit margin} for scenario 3 is set to 200\%, meaning the data provider's total earnings in this scenario are exactly double that of their costs (making for 100\% pure profit after covering costs). 


\begin{table}
\caption{Base cost for the core functions of LUCE.}
\label{table:basecost}
\setlength{\tabcolsep}{3pt}
\def\arraystretch{1.5}
\begin{tabular}{p{110pt}p{60pt}p{60pt}p{60pt}p{60pt}}
 Actions & Transaction cost & Execution cost &  Ether cost & Cost* \\\hline 
\textit{Deployment} & 6724230 & 5118378 & 0.48414 & \$831.03 \\ 
\hline
\textit{publishData} & 95560 & 72560 & 0.00688 & \$11.80 \\ 
\hline
\textit{updateData} & 43799 & 20863 & 0.00315 & \$5.40 \\ 
\hline
\textit{addDataRequester} & 475067 & 453411 & 0.03420 & \$58.70 \\ \hline
\textit{renewToken} & 45211 & 23747 & 0.00326 & \$5.59 \\ \hline
\textit{setLicense} & 39339 & 37075 & 0.00283 & \$4.85 \\ \hline
\textit{setRegistryAddress} & 37131 & 14515 & 0.00267 & \$4.58 \\ \hline
\textit{setProfitMargin} & 35091 & 13627 & 0.00253 & \$4.34 \\ \hline
\textit{setPrice} & 31062 & 9406 & 0.00224 & \$3.84 \\ \hline
\multicolumn{5}{p{350pt}}{*= Ether conversion with present date price}
\end{tabular}
\end{table}
\subsection{Cost analysis} 
\label{Cost analysis}
\begin{figure}[ht]
\centering
\includegraphics[width=\textwidth]{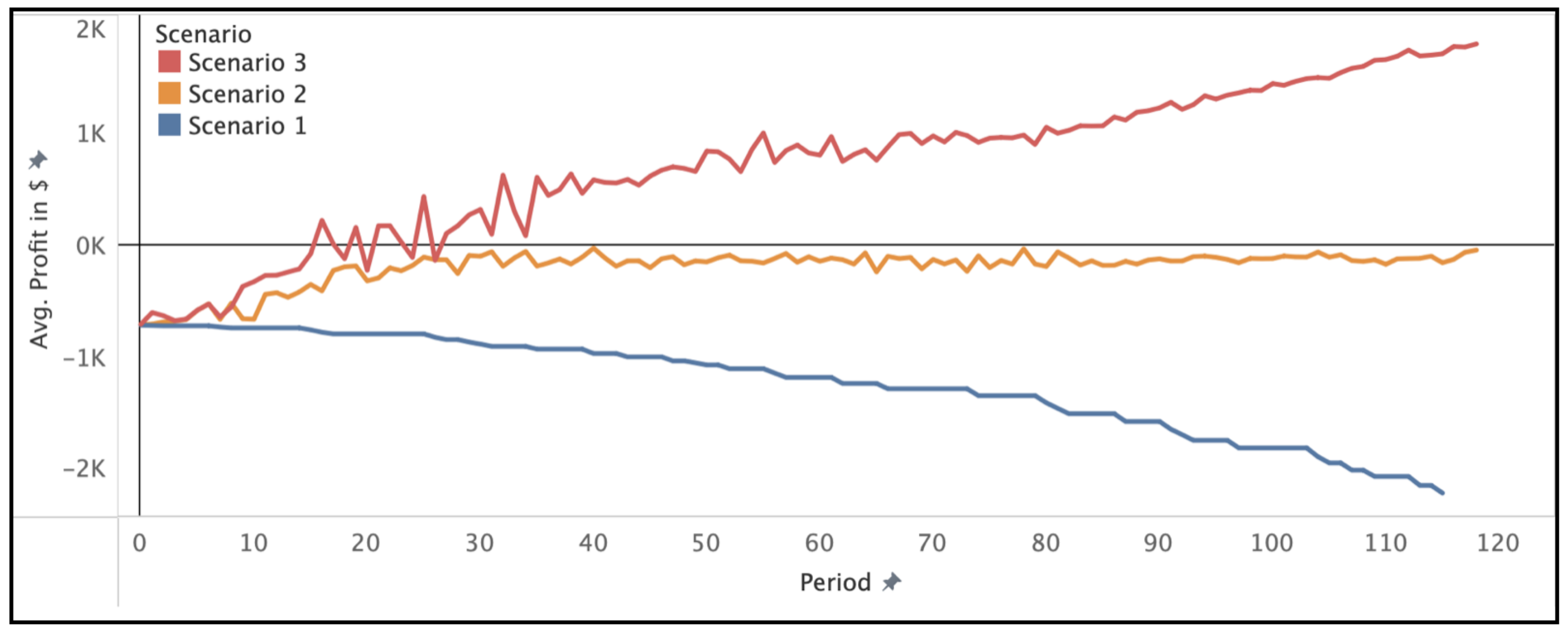}
\caption[Profit per scenario]{Profit over time for each scenario.}
\label{fig:Profit per scenario}
\end{figure}

Transactions on the Ethereum network have a \textit{gas} cost that is directly proportional to the internal operations of the respective function call in the smart contract. Specifically, storing data on the blockchain is relatively expensive, therefore, the cost of writing to the blockchain scales with the size of the content. Thus, the deployment cost of a new smart contract is generally quite high compared to transactions resulting from calling the functions of that smart contract. 
Table~\ref{table:costparameters} describes the cost parameters used in the incentive model. Table~\ref{table:basecost} shows the base costs of the core functions of LUCE whereas ~\ref{table:costregistry} shows the cost of the core functions of the LUCE registry smart contract. These are the pure transaction costs resulting from calling the respective function, which equates to scenario 1. In scenario 2 and 3, the \textit{request} and \textit{renew} functions require additional funds to be transmitted with each function call. As mentioned before, the costs to update a dataset scale with its active users. Therefore the cost is relatively low when there is no data requester (\$5.40), and far higher when there are e.g. 60 data requesters (\$64.30), which makes for roughly \$1.07 per requester for an update. Figure~\ref{fig:Profit per scenario} shows that these comparatively higher costs are still easily covered by the system. It shows the profits generated in each scenario. We can see that after approximately 40 periods in scenario 2, costs are completely covered, whereas, in scenario 3 the break-even point is reached faster, and positive returns are measured as soon as period 16.

\begin{figure}[ht]
\centering
\includegraphics[width=\linewidth]{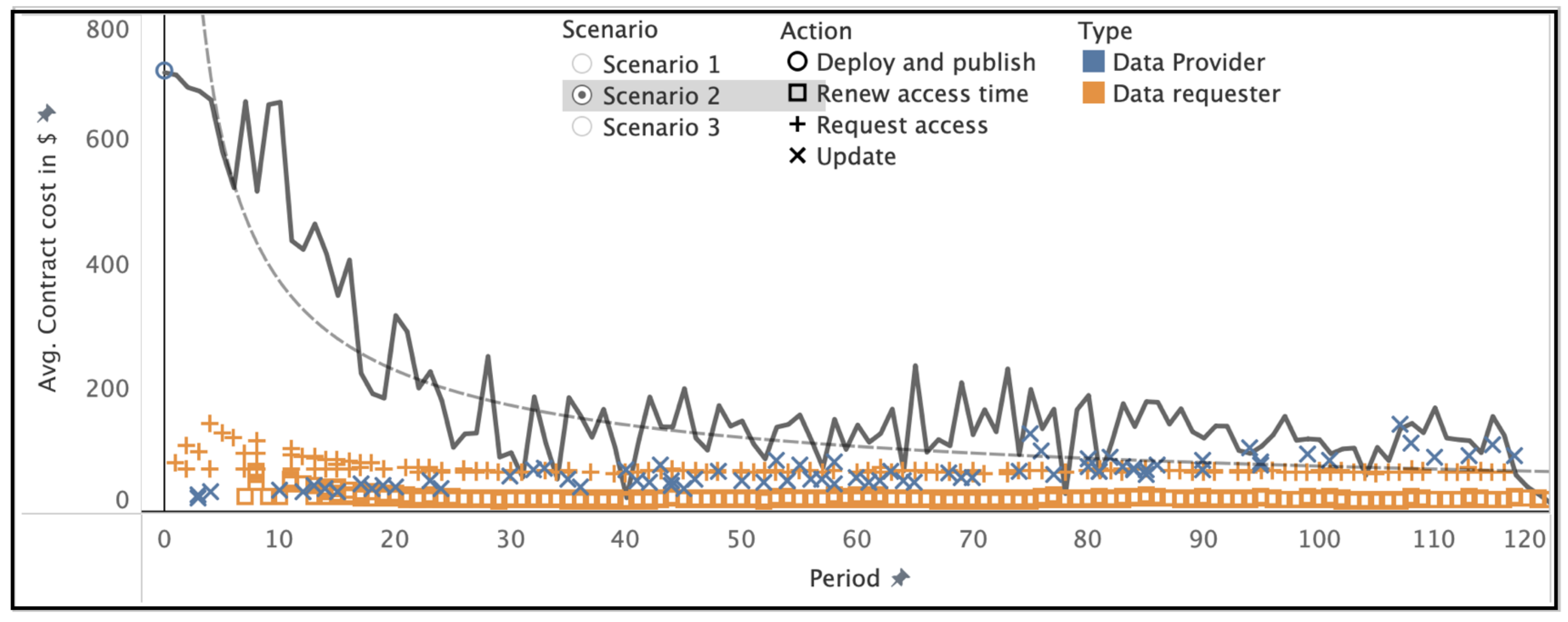}
\caption[Scenario 2: Cost and Transactions]{Running costs of a smart contract over a period of time in Scenario 2.}
\label{fig:Scenario 2 Cost and Transactions}
\end{figure}
The cost of updating the meta-information of the data in the smart contract scales with the number of requesters since each requester must be notified of that update to give them a chance to comply. Figure~\ref{fig:Scenario 2 Cost and Transactions} displays the relationship of running contract costs (grey line; the spikes are updates) and individual transactions in more detail. We observe that the running costs of a smart contract are influenced by individual transactions made by the data provider and data requesters. Here, we can more closely observe rising update costs (the blue X marks) and sinking access costs over time (the orange squares and plus signs). Each data requester in this scenario pays 5\% of the running costs at the time of their request. With this setting, data providers in scenario 2 can veritably expect that their costs will always be covered under the condition that data requesters continue to use their dataset. If the dataset loses its value, cost coverage may take a longer time, or, in extreme cases, costs may not be covered. In our simulation, the only difference between scenario 2 and scenario 3 is the \textit{profit margin}. Profits in scenario 3 are effectively a linear multiplication of costs in scenario 2 and follow the same arguments. However, since scenario 3 is explicitly profitable, it reaches the break-even point faster in proportion to how high the \textit{profit margin} is set. 
\begin{table}
\centering
\caption{Cost listing of all functions of the LUCE Registry.}
\label{table:costregistry}
\setlength{\tabcolsep}{3pt}
\def\arraystretch{1.5}
\begin{tabular}{p{110pt}p{60pt}p{60pt}p{60pt}p{60pt}}
 Actions & Transaction cost & Execution cost &  Ether cost & Cost* \\\hline 
\textit{Deployment} & 621087 & 432315 & 0.04472 & \$76.76 \\ 
\textit{newDataProvider} & 44855 & 22175 & 0.00323 & \$5.54 \\ 
\textit{registerNewUser} & 45669 & 22797 & 0.00329 & \$5.64 \\ 
\textit{updateUserLicense} & 27732 & 6268 & 0.00200 & \$3.43 \\
\textit{checkProvider} & 23991 & 1311 & 0.00173 & \$2.96 \\ 
\textit{checkUser} & 23877 & 1197 & 0.00172 & \$2.95 \\ \hline
\multicolumn{5}{p{350pt}}{*= Ether conversion with present date price}
\end{tabular}
\end{table}
We can also observe the change in additional costs for data requesters. After initial deployment (periods 1-20), costs for requesters are higher than otherwise (periods after 20). In figure \ref{fig:Scenario 2 Cost and Transactions}, there are 59 data requesters in total, simulated over 118 periods. Specifically, 27 updates to the data (frequency 0.22/period), 59 access requests (frequency 0.48/period), and 418 access time renewals (frequency 3.54/period). This makes a total of 505 \textit{actions} and reflects our assumption that there are far more data requesters than providers. The initial cost for a data requester is dependent on which scenario we are simulating. As mentioned in table \ref{table:basecost}, the base cost of requesting access is \$58.70. In the other two scenarios, a variable additional price is added to cover the data provider's cost or generate the data provider's profit respectively. 

\begin{figure}[ht]
\centering
\includegraphics[width=0.65\linewidth]{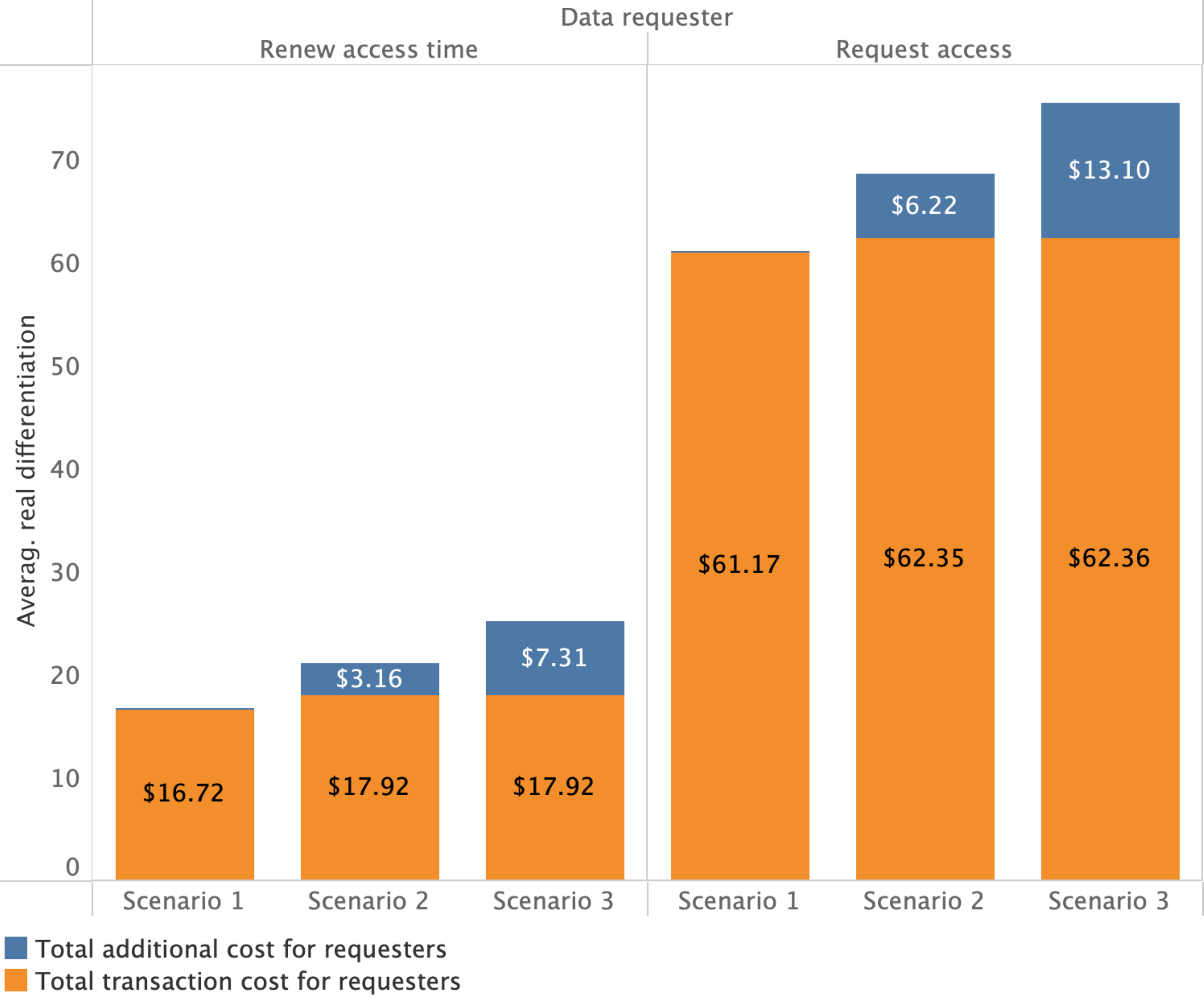}
\caption[Cost breakdown for data requesters]{Additional transaction costs for data requesters to access a dataset.}
\label{fig:Cost breakdown}
\end{figure}

\begin{figure}[ht]
\centering
\includegraphics[width=0.65\linewidth]{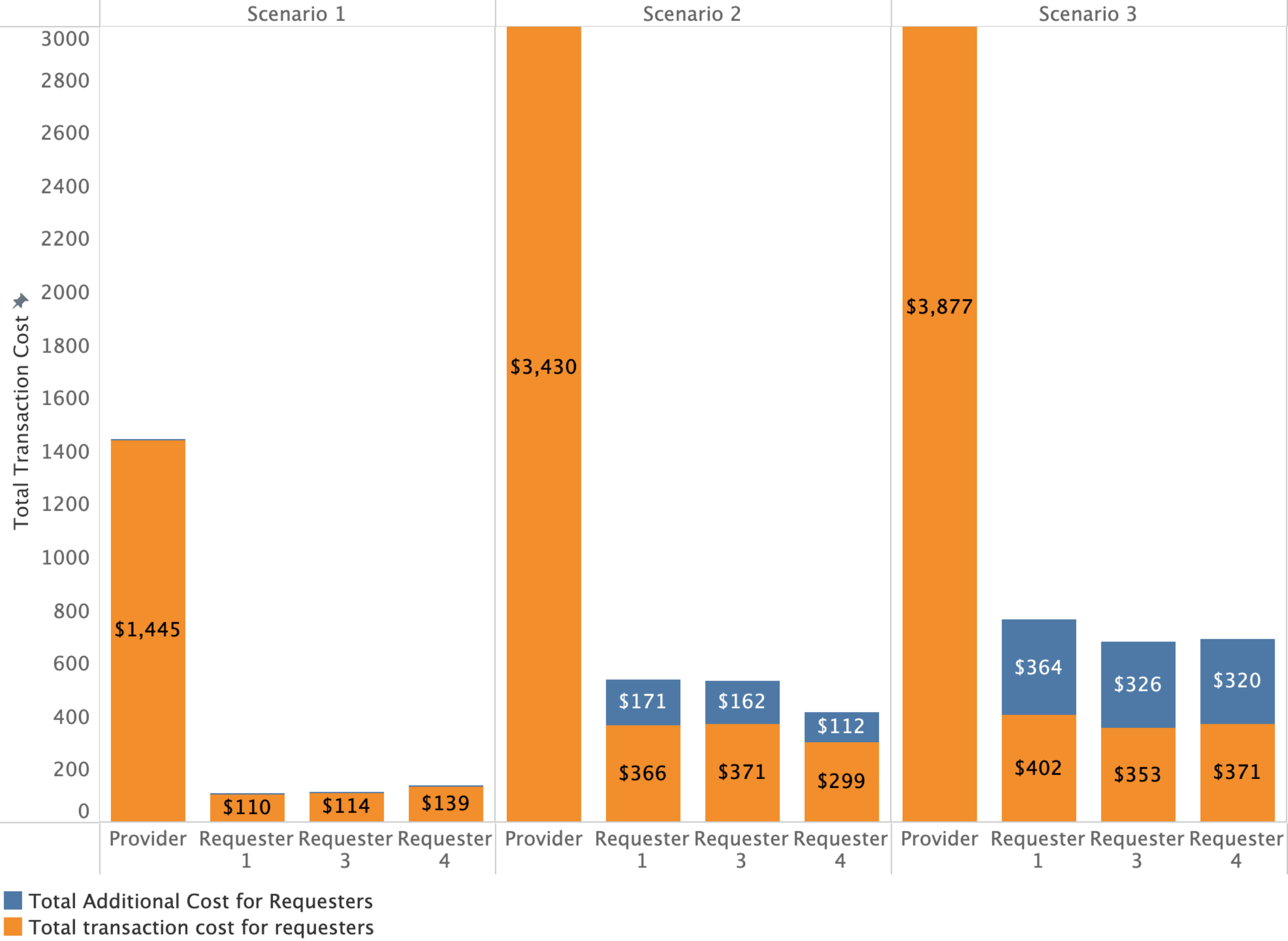}

\caption[Total Cost of Provider and Requesters per Scenario]{Total cost sum of the data provider vs top three data requesters per scenario.}
\label{fig:TopRequesters}
\end{figure}

\begin{figure}[ht]
\centering
\includegraphics[width=0.65\textwidth]{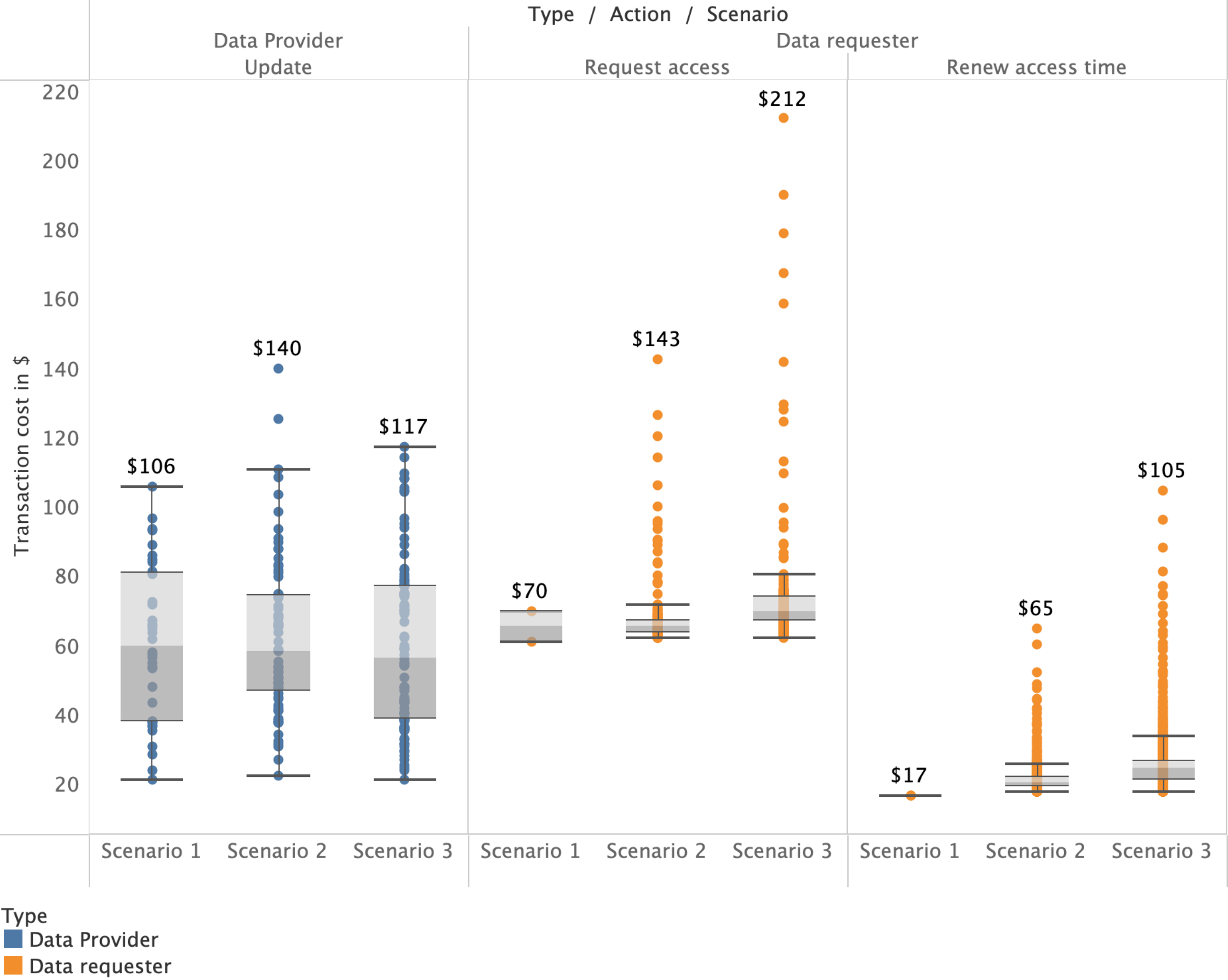}
\caption[Cost per action and user type]{Range of costs arising for each user and \textit{action} type.}
\label{fig:Cost per action and user type}
\end{figure}

Figures \ref{fig:Cost breakdown} and \ref{fig:TopRequesters} show requester costs specific to each scenario. We can observe the average base transaction cost for requester \textit{action} types and the additional cost stacked on top (which the requester bears instead of the provider in the case of scenarios 2 or 3 respectively). Compared to requesters' individual costs, the data provider has much higher costs, as shown in figure \ref{fig:TopRequesters}. Over 118 periods, data providers must invest between \$1445 to \$3877. However, as demonstrated by our simulations, even the relatively high initial costs of deployment can be expected to be quickly recovered by the data provider in the case of scenarios 2 and 3. This reflects the assumption that there are far more data requesters than providers. If this were not the case, data providers would likely be forced to set higher cost allocation fractions to cover their costs. 
For a more detailed overview of what range of costs each user of the platform can expect, we plot the simulated cost distributions based on each \textit{action} type in figure \ref{fig:Cost per action and user type} in a logarithmic manner. 
We can infer from figure \ref{fig:Cost per action and user type} that there are few outliers concerning the cost distribution among data requesters, both when initially requesting access, and when renewing that access. This can be attributed to the fact that the first five to ten early requesters cover the majority of initial deployment costs, which are generally much higher than update costs. This unequal distribution of costs could be smoothed out by choosing a smaller fraction to denote the percentage of the running costs requesters must cover. If this fraction is chosen too small, it will likely lead to the data provider's cost not being covered, which defeats the purpose of scenario 2.

With the simulations of the three distinct scenarios, we show that depending on the parameters set in the smart contract, (a) data providers face considerable up-front costs to cover the deployment of pertinent smart contracts, and (b) the initial investment, as well as all running costs incurred by the necessity to regularly update the data, can be rapidly recovered by data providers. Importantly, this suggests that both \textit{cost} and \textit{monetary} incentives would likely be effective means to motivate data providers to participate in the LUCE platform. In scenario 2, the ability for data providers to quickly recover up-front investments minimizes the disincentive that up-front costs would otherwise manifest. Consequently, the main positive incentives in scenario 2, \textit{pertinent} and \textit{reputation}, will likely not be significantly diminished by cost. Scenario 3 extends this by additionally introducing a \textit{monetary} incentive. Here, costs incurred by data providers are covered with even stronger kinetics than in scenario 2, and they additionally benefit from profits, determined by the \textit{profit margin} they set. Through our implementation of how payments by data requesters are calculated, data providers effectively cannot profit infinitely, depending on the frequency with which they make updates to their data. The profit calculation is directly derived from occurring costs. Therefore, if data requesters sufficiently outnumber data providers, there will come a point where the data provider has fully achieved their desired profit because it is a linear combination of their costs. From that point, only new costs incurred by the data provider, e.g. an update to their data, will result in net profit. This effectively limits how much a data provider can ever profit from high demand and since the same calculation is used for scenario 2, where no profit is generated, high demand will similarly result in costs being covered completely, which means requesters have no additional costs from requesting access to the data. In such cases of extremely high demand, it may be a valid fairness consideration of the data provider to lower the percentage of the running costs each data requester must pay. Conversely, if there is extremely low demand, the data provider may wish to increase this percentage. As such, we provide the data provider the tools they need to control how their costs are covered or profits are generated. 
\section{Discussion}\label{sec:discussion}
\subsection{Incentives}
Our results show that in scenario 2 the costs of the data provider are quickly recovered. An important question that remains is how long this will take in the real world? This time should not be unreasonably high. If we assume that one period equates to one week, then complete cost coverage will take approximately seven months. Conversely, if we assume that a period is a day, it will take less than one month to cover all costs. However, since this is based on stringent assumptions about the users of the system, it is impossible to deduce the number that reflects reality. The only way to reasonably predict this will be a study that surveys how data subjects, providers, and requesters would act if they had access to the system. Nevertheless, given the low relative costs of data provision for the presumed participants, even a conservative estimate of cost-recovery over several months would likely not present a significant disincentive for data providers.
\subsection{Costs}
Additionally, we do not consider costs resulting from the ETL processes required to make data useful for analysis. Large data providers (i.e a medical center) may need to employ people to facilitate the compilation of relevant data to be shared on the LUCE platform. These costs could be injected into the smart contract logic, and data requesters will ultimately defray these additional costs. However, if our assumption holds that data requesters far outnumber providers, this additional cost will likely not outstrip the costs by an insurmountable margin. 
\section{Related Work}\label{sec:relatedwork}
Several works have focused on data-sharing incentives for decentralized networks. Shrestha et al.~\cite{DataSharingFramework} introduce a basic functioning framework for data-sharing via blockchain authentication. Apart from the system's inherent data-sharing incentives, authors focus on a monetary compensation incentive for data providers. The authors, however, do not show the specifications of incentive mechanisms of requesters to providers i.e. whether profit is generated or the system strives to achieve break-even. 
In this paper, we contribute a detailed perspective of costs resulting from data-sharing platforms utilizing a comprehensive, extended, and easily reproducible prototype with sophisticated smart contract logic. We show how users can be incentivized to participate in the platform, and what ramifications different cost allocations result in the system. 

The Ocean protocol~\cite{OceanWhitePaper} functions as a Marketplace listing all available datasets. Data providers hold the data themselves and only release it when there is a legitimate request, verifiable through a respective entry in the underlying blockchain smart contract. 
The economy of Ocean is based on their in-house crypto-token called OCN. 
The OCN token discourages sharing poor quality data by implementing a staking mechanism that ties the provided data to personal assets - high-quality data would then result in reaching the break-even point quickly~\cite{OceanMarketplace}. The drawback is to use of their in-house token adds a layer of complication to the system that does not necessarily ensure asset value-retention, since Ocean actively avoided implementation of price stability due to performance concerns. Another drawback is the lack of autonomous tools for the data provider and data subject to directly, effectively facilitate GDPR compliance~\cite{GDPR}. We present LUCE with monetary-based incentive mechanisms which are GDPR compliant and data owners can set their terms in the smart contract logic while sharing the data.  
Xuan et al.~\cite{IncentiveStrategy} offer a mathematical analysis of participation strategies in blockchain-based data-sharing applications based on game theory. Authors derive four conditions for which they model user participation in the system and create an incentive method that results in a stable user base, i.e. no over or undersaturation of users willing to share data. This could provide a basis for a more sophisticated simulation that derives participation probabilities from gain functions and pricing strategies. However, the authors do not detail the data requesters' payment structures to pay for the data or the consequences for the bad quality of data received by them. Our incentive-based approach gives a balanced view of the system with different incentive strategies and is GDPR compliant. Reputation-based approaches~\cite{blockchainreputation,blockchainreputation2} have also been proposed where service providers and requesters are not supposed to be trusted. Service requesters use reputation-based credentials to choose the service providers which is a perception of the service provider's past behavior. Privacy-preserving incentive mechanisms~\cite{incentiveprivacy,reportcoin} such as ReportCoin~\cite{reportcoin} where it motivates users to publish anonymous reporting and incentive is received via their \textit{Rcoins}. However, in this paper, we only consider and simulate the monetary-based compensation. Some other approaches~\cite{incentivesShen2020,incentiveIoT} include incentive mechanisms for data sharing in IoT and clouds. The authors' approaches include the Shapley value, which is commonly used for resource sharing and revenue distribution models. However, the authors also raised the challenge of achieving a fair distribution of benefits. In our future work, we will test application in a closed environment with real participants to understand the behavior towards the system and how incentives contribute to it.

\section{Conclusion and Future work}\label{sec:conclusion}
In this paper, we present incentive mechanisms for blockchain-based data sharing platforms. We propose multiple smart contracts that dynamically adjust incentives and participation costs.
Using multiple cost pricing scenarios for data owners we simulate data monetization strategies. We conclude that a cost compensation incentive model can rapidly cover the cost of data sharing, thus encouraging data owners to share data in the platform. 
In the future, we will study end-user interactions to best understand other forms of incentives, such as knowledge sharing, and how that may impact the dynamics in a data-sharing network. We will also further explore other monetization strategies and generate more sophisticated simulations that derive participation probabilities from pricing strategies.  
\section*{Acknowledgment}
This work was supported in part by the NWO Aspasia (Grant 91716421) and by the Maastricht York Partnership Grant.

\bibliographystyle{unsrt}  
\bibliography{bibliography}  


\end{document}